\documentclass{article}
\usepackage[table]{xcolor} 
\usepackage{tabu}
\usepackage{spconf,amsmath,graphicx}
\usepackage{booktabs}
\usepackage{tabularx}
\usepackage{url}
\usepackage{multirow,bigdelim}
\usepackage{pbox}
\usepackage{svg}
\usepackage{hyperref}
\usepackage{tabu}
\usepackage{multirow}
\usepackage{bigdelim}

\newcolumntype{a}{>{\columncolor{blue!10}}c}
\newcolumntype{x}{>{\columncolor{blue!20}}c}

\hypersetup{
    colorlinks=true,
    citecolor=blue,
    linkcolor=blue,
    filecolor=magenta,      
    urlcolor=blue,
}

\title{THE ICASSP 2024 AUDIO DEEP PACKET LOSS CONCEALMENT GRAND CHALLENGE} 
\name{Lorenz Diener, Solomiya Branets, Ando Saabas, Ross Cutler}
\address{Microsoft Corp.}

\begin{document}

\maketitle

\begin{abstract}
Audio packet loss concealment is the hiding of gaps in VoIP audio streams caused by network packet loss. With the ICASSP 2024 Audio Deep Packet Loss Concealment Grand Challenge, we build on the success of the previous Audio PLC Challenge held at INTERSPEECH 2022. We evaluate models on an overall harder dataset, and use the new ITU-T P.804 evaluation procedure to more closely evaluate the performance of systems specifically on the PLC task.

We evaluate a total of 9 systems, 8 of which satisfy the strict real-time performance requirements of the challenge, using both P.804 and Word Accuracy evaluations.
\end{abstract}
\section{Introduction}

As voice communication further transitions more and more towards calls that are fully packet switched end-to-end (rather than being fully circuit switched, or circuit switched with a dedicated packet switched backbone), the need for more robust packet loss concealment has never been more evident. Since the tight latency requirements in real-time communication applications make large buffers and retransmission undesirable if not impossible, degraded network performance leads to audible gaps or annoying distortion in calls at the receiver side. Audio Packet Loss Concealment (PLC) is the task of fixing or hiding these gaps and making the audio stream appear as seamless as possible to allow for high quality communication even when packets get lost.

As algorithms and hardware have advanced, it is now possible to perform PLC using machine learning rather than basic digital signal processing, with potential for vast quality improvements. In the PLC Challenge held at INTERSPEECH 2022, we for the first time brought together researchers working on the topic to compare approaches on a common dataset, with many interesting approaches and results~\cite{diener2022interspeech}.

\section{Changes from the 2022 PLC Challenge}
In this edition of the challenge, we build on this success, and make some changes based on lessons learned:

\textbf{A more challenging dataset:} The dataset in the 2022 PLC Challenge was, while not easy, still largely focused scenarios where there is only a relatively low amount of packets lost, with not too many packets being lost in a row. Many participants built systems with good performance in these scenarios, but which may not be able to perform well for longer sequences of losses. To challenge participants to also tackle harder cases with long burst losses, the dataset in this challenge focuses more on such cases. Additionally, while the 2022 challenge used wideband audio, the audio used in this edition is full-band, making the task once again somewhat more difficult, especially given the latency and compute constraints remain unchanged.

\textbf{Better evaluation procedure:} In the 2022 challenge, we performed objective evaluation using a ITU-T P.808 CCR procedure, obtaining a single rating for each file. In this challenge, we switch to the newer ITU-T P.804~\cite{naderi2023multi} standard, in which listeners are asked to evaluate an audio file on multiple scales: Coloration, Noisiness, Discontinuity, Reverb, Signal Quality and Overall Quality.

\begin{table*}[h!t]
\extrarowheight=\aboverulesep
\addtolength{\extrarowheight}{\belowrulesep}
\aboverulesep=0pt
\belowrulesep=0pt
\centering
\caption{ICASSP 2024 Audio Deep PLC Challenge results. Differences between systems significant at $p < 0.05$ except where indicated (NS). Columns that contributed to final evaluation score in the challenge highlighted, best system for every metric bolded.}
\label{tab:challenge-results}
\setlength{\tabcolsep}{5pt} 
\begin{tabular}{@{}llccaccaacc@{}}
\toprule
          & \multicolumn{6}{c}{\hspace{140px}P.804 Scores} &   \multicolumn{2}{c}{}              \\
\cmidrule(l){3-8}
Place & System & Coloration & Noisiness & \multicolumn{1}{c}{Discontinuity} & Reverb & Signal & \multicolumn{1}{c}{Overall} & \multicolumn{1}{c}{WAcc} & Final Score  \\
\midrule
           &       \color{gray}Raw (Clean) &       \color{gray}4.43 &  \color{gray}4.17 &          \color{gray}4.55 &   \color{gray}4.40 &   \color{gray}4.34 &    \color{gray}4.01 &          \color{gray}0.98 &    \color{gray}0.87 \\
1 (shared) &         1024K &       4.05 &  \textbf{4.26} &          \textbf{3.90} &   \textbf{4.21} &   \textbf{3.78} &    \textbf{3.49} &          0.81 &     0.72 & \hspace{-20px}\rdelim\}{2}{4mm}[NS] \\
1 (shared) &       NWPU \& ByteAudio &       \textbf{4.11} &  4.03 &          3.82 &   4.14 &   3.73 &    3.44 &          \textbf{0.84} &    0.72 \\
3          & SpeechGroupIoA &       4.05 &  4.23 &          3.67 &   4.15 &   3.64 &    3.37 &          0.81 &    0.69 \\
4          &         HWYW &       3.99 &  4.14 &          3.49 &   4.09 &   3.49 &    3.21 &          0.81 &    0.66 \\
5          &        LEIBUS &       3.74 &  3.82 &          2.94 &   3.87 &   2.98 &    2.75 &          0.84 &    0.59 \\
6          &   Regenerate &       3.53 &  3.42 &          2.90 &   3.64 &   2.83 &    2.56 &          0.83 &    0.57 \\
           &  \color{gray}Raw (Lossy) &       \color{gray}3.60 &  \color{gray}3.67 &          \color{gray}2.47 &   \color{gray}3.72 &   \color{gray}2.58 &    \color{gray}2.37 &          \color{gray}0.83 &   \color{gray}0.51 \\
7          & CQUPT\_ISARL &       2.93 &  3.19 &          2.65 &   3.13 &   2.34 &    2.11 &          0.81 &   0.50 \\
8          &   NJUAcstcs &       2.92 &  3.18 &          2.68 &   3.15 &   2.39 &    2.17 &          0.64 &   0.45 \\
(DNF)      & Enchanto &       3.73 &  3.59 &          3.36 &   3.85 &   3.21 &    2.91 &          0.82 &   0.63 \\
\bottomrule
\end{tabular}
\vspace{-0.8em}
\end{table*} 

\section{Challenge description}
\subsection{Dataset construction}
The dataset for the ICASSP 2024 challenge is built upon the same framework as before, leveraging real-world packet loss patterns combined with data that is either in the public domain (conversational speech, sourced from the LibriVox Community Podcast)\footnote{Librivox Contributors, ``The LibriVox community podcast'', \url{https://librivox.org/category/librivox-community-podcast/}} or was collected by us explicitly for use in challenges (read speech), allowing us to have a realistic dataset while avoiding the potential for privacy issues. Audio segments were selected by filtering using DNSMOS~\cite{reddy2021dnsmos} and manual inspection to avoid very noisy base audio clips, and were cut to 10 to 15 seconds of length using the WebRTC Voice Activity Detection to avoid cutting off parts of words. All clips were normalized to $-6~dBFS$ peak amplitude.

Packet loss traces were selected as follows: First, we select packet loss trace segments according to the longest burst loss present (exclusive higher edge). We then select traces for 5 equally sized packet loss brackets (0\% to 10\%, 10\% to 20\%, 20\% to 30\%, 30\% to 40\%, above 40\%) for each of these burst loss ranges. We select a total of 600 traces:

\begin{itemize}\setlength\itemsep{-0.2em}
    \item \textbf{0--120~ms burst:} 20 per loss bracket, 100 total
    \item \textbf{120--500~ms burst:} 40 per loss bracket, 200 total
    \item \textbf{500--1000~ms burst:} 40 per loss bracket, 200 total
    \item \textbf{1000--3000~ms burst:} 20 per loss bracket, 100 total
\end{itemize}

Additionally, we include a total of 200 traces also used in the 2022 PLC Challenge to allow for limited comparability. Further details about the dataset creation procedure and data sourcing can be found in our previous work~\cite{diener2022interspeech}. On Oct.~11, 2023, we first released a validation set constructed in this way, followed by a blind set with no references on Dec.~1, 2023. Participants submitted the output of their systems for this blind set for evaluation  by the deadline of Dec.~7, 2023.

\subsection{Evaluation procedure}
We perform a P.804 evaluation using the Amazon Mechanical Turk crowd-sourcing service. For quality control, we include both two gold questions (clips where the expected answer for a scale is known ahead of time, with either very low or very high quality) and one trapping question (questions where the rating clip is replaced by instructions to select a specific answer regardless of quality). We only use answers from listeners that consistently answer these quality control questions correctly~\cite{naderi2023multi}. After quality filtering, we obtain on average approx. 5 ratings for each clip.

\section{Results and Conclusion}
The results can be found in Table~\ref{tab:challenge-results}. We also include one system that did not meet latency requirements (marked DNF). We perform statistical testing (one-tailed related-sample t-test between systems adjacent to each other in the scoreboard, no FWER correction) on the final score to see whether the differences we obtain are significant. Based on this, the winners of the ICASSP 2024 Audio Deep Packet Loss Concealment Grand Challenge are, sharing the first place, teams 1024K and NWPU \& ByteAudio. For detailed results, including objective metrics, and copies of both validation and blind set with reference audio included, please refer to our challenge website~\footnote{\url{https://aka.ms/plc_challenge}}. For details on the top 5 systems from the challenge, please refer to participant challenge papers.

We would like to thank all participants for their submissions, and hope that the challenge has served to move the state of the field of machine learning based packet loss concealment forward.

\bibliographystyle{IEEEbib}
\bibliography{main}

\begin{thebibliography}{1}

\bibitem{diener2022interspeech}
Lorenz Diener, Sten Sootla, Solomiya Branets, Ando Saabas, Robert Aichner, and Ross Cutler,
\newblock ``{INTERSPEECH 2022 Audio Deep Packet Loss Concealment Challenge},''
\newblock in {\em Proc. Interspeech 2022}, 2022, pp. 580--584.

\bibitem{naderi2023multi}
Babak Naderi, Ross Cutler, and Nicolae-Catalin Ristea,
\newblock ``Multi-dimensional speech quality assessment in crowdsourcing,''
\newblock in {\em ICASSP}, 2024.

\bibitem{reddy2021dnsmos}
Chandan~KA Reddy, Vishak Gopal, and Ross Cutler,
\newblock ``{DNSMOS}: A non-intrusive perceptual objective speech quality metric to evaluate noise suppressors,''
\newblock in {\em ICASSP}, 2021.

\end{thebibliography}

\end{document}